\documentclass{sigkddExp}
\usepackage{multirow,colortbl,tablefootnote,graphicx,epstopdf,color,booktabs,amsmath,diagbox}
\usepackage{balance}
\usepackage{hyperref}
\hypersetup{
    colorlinks=true,
    linkcolor={black},
    citecolor={blue},
    urlcolor={black}
}
\usepackage[table]{xcolor}
\usepackage[flushleft]{threeparttable}
\usepackage[font=footnotesize]{subcaption}

\newlength{\bibitemsep}
\newlength{\bibparskip}
\let\oldthebibliography\thebibliography
\renewcommand\thebibliography[1]{%
  \oldthebibliography{#1}%
  \setlength{\parskip}{\bibitemsep}%
  \setlength{\itemsep}{\bibparskip}%
}

\begin{document}

\title{Fake News Detection on Social Media: \\ A Data Mining Perspective}

\numberofauthors{1}
\author{
\alignauthor Kai Shu$^\dagger$, Amy Sliva$^\ddagger$,  Suhang Wang$^\dagger$, Jiliang Tang $^\natural$, and Huan Liu$^\dagger$\\
		$^\dagger$\affaddr{Computer Science \& Engineering, Arizona State University, Tempe, AZ, USA}\\
		$^\ddagger$\affaddr{Charles River Analytics, Cambridge, MA, USA}\\
 $^\natural$\affaddr{Computer Science \& Engineering, Michigan State University, East Lansing, MI, USA}
		\email{$^\dagger$\{kai.shu,suhang.wang,huan.liu\}@asu.edu,  \\ $^\ddagger$asliva@cra.com, $^\natural$ tangjili@msu.edu}
}

\maketitle
\begin{abstract}

Social media for news consumption is a double-edged sword. On the one hand, its low cost, easy access, and rapid dissemination of information lead people to seek out and consume news from social media. On the other hand,  it enables the wide spread of ``fake news'', i.e., low quality news with intentionally false information. The extensive spread of fake news has the potential for extremely negative impacts on individuals and society. Therefore, fake news detection on social media has recently become an emerging research that is attracting tremendous attention. Fake news detection on social media presents unique characteristics and challenges that  make existing detection algorithms from traditional news media  ineffective or not applicable. First, fake news is intentionally written to  mislead readers to believe false information, which makes it difficult and nontrivial to detect  based on news content; therefore, we need to include auxiliary information, such as user social engagements on social media, to help make a determination. Second, exploiting this auxiliary information is challenging in and of itself as users' social engagements with fake news produce data that is big, incomplete, unstructured, and noisy. Because the issue of fake news detection on social media is both challenging and relevant, we conducted this survey to further facilitate research on the problem. In this survey, we present a comprehensive review of detecting fake news on social media, including fake news characterizations on psychology and social theories, existing algorithms from a data mining perspective, evaluation metrics and representative datasets. We also discuss related research areas, open problems, and future research directions for fake news detection on social media.


\end{abstract}
\section{Introduction}

As an increasing amount of our lives is spent interacting online through social media platforms, more and more people tend to seek out and consume news from social media rather than traditional news organizations. The reasons for this change in consumption behaviors are inherent in the nature of these social media platforms: (i) it is often more timely and less expensive to consume news on social media compared with traditional news media, such as newspapers or television; and (ii) it is easier to further share, comment on, and discuss the news with friends or other readers on social media. For example, 62 percent of U.S. adults get news on social media in 2016, while in 2012, only 49 percent reported seeing news on social media\footnote{http://www.journalism.org/2016/05/26/news-use-across-social-media-platforms-2016/}. It was also found that social media now outperforms television as the major news source\footnote{http://www.bbc.com/news/uk-36528256}. Despite the advantages provided by social media, the quality of news on social media is lower than traditional news organizations. However, because it is cheap to provide news online and much faster and easier to disseminate through social  media, large volumes of fake news, i.e.,  those news articles with intentionally false information, are produced online for a variety of purposes, such as financial and political gain.  It was estimated that over 1 million tweets are related to fake news ``Pizzagate''\footnote{https://en.wikipedia.org/wiki/Pizzagate\_conspiracy\_theory} by the end of the presidential election. Given the prevalence of this new phenomenon, ``Fake news'' was even named the word of the year by the Macquarie dictionary in 2016.

The extensive spread of fake news can have a serious negative impact on individuals and society. First, fake news can break the authenticity balance of the news ecosystem. For example, it is evident that the most popular fake news was even more widely spread on Facebook than the most popular authentic mainstream news during the U.S. 2016 president election\footnote{https://www.buzzfeed.com/craigsilverman/viral-fake-election-news-outperformed-real-news-on-facebook?utm\_term=.nrg0WA1VP0\#.gjJyKapW5y}. Second, fake news intentionally persuades consumers to accept biased or false beliefs. Fake news is usually manipulated by propagandists to convey political messages or influence. For example, some report shows that Russia has created fake accounts and social bots to spread false stories\footnote{http://time.com/4783932/inside-russia-social-media-war-america/}. Third, fake news changes the way people interpret and respond to real news. For example, some fake news was just created to trigger people's distrust and make them confused, impeding their abilities to differentiate what is true from what is not\footnote{https://www.nytimes.com/2016/11/28/opinion/fake-news-and-the-internet-shell-game.html?\_r=0}. To help mitigate the negative effects caused by fake news--both to benefit the public and the news ecosystem--It's critical that we develop methods to automatically detect fake news on social media.

Detecting fake news on social media poses several new and challenging research problems. Though fake news itself is not a new problem--nations or groups have been using the news media to execute propaganda or influence operations for centuries--the rise of web-generated news on social media makes fake news a more powerful force that challenges traditional journalistic norms. There are several characteristics of this problem that make it uniquely challenging for automated detection. First, fake news is intentionally written to mislead readers, which makes it nontrivial to detect simply based on news content. The content of fake news is rather diverse in terms of topics, styles and media platforms, and fake news attempts to distort truth with diverse linguistic styles while simultaneously mocking true news. For example, fake news may cite true evidence within the incorrect context to support a non-factual claim~\cite{feng2012syntactic}. Thus, existing hand-crafted and data-specific textual features are generally not sufficient for fake news detection. Other auxiliary information must also be applied to improve detection, such as knowledge base and user social engagements. Second, exploiting this auxiliary information actually leads to another critical challenge: the quality of the data itself. Fake news is usually related to newly emerging, time-critical events, which may not have been properly verified by existing knowledge bases due to the lack of corroborating evidence or claims. In addition, users' social engagements with fake news produce data that is big, incomplete, unstructured, and noisy~\cite{tang2014mining}. Effective methods to differentiate credible users, extract useful post features and exploit network interactions are an open area of research and need further investigations. 


In this article, we present an overview of fake news detection and discuss promising research directions. The key motivations of this survey are summarized as follows:
\begin{itemize}
\item Fake news on social media has been occurring for several years; however, there is no agreed upon definition of the term ``fake news''. To better guide the future directions of fake news detection research, appropriate clarifications are necessary. 
\item Social media has proved to be  a powerful source for fake news dissemination. There are some emerging patterns that can be utilized for fake news detection in social media. A review on existing fake news detection methods under various social media scenarios can provide a basic understanding on the state-of-the-art fake news detection methods.
\item Fake news detection on social media is still in the early age of development, and there are still many challenging issues that need further investigations. It is necessary to discuss potential research directions that can improve fake news detection and mitigation capabilities.
\end{itemize}

\begin{figure*}[tb!]
\centering
\includegraphics[scale=0.45]{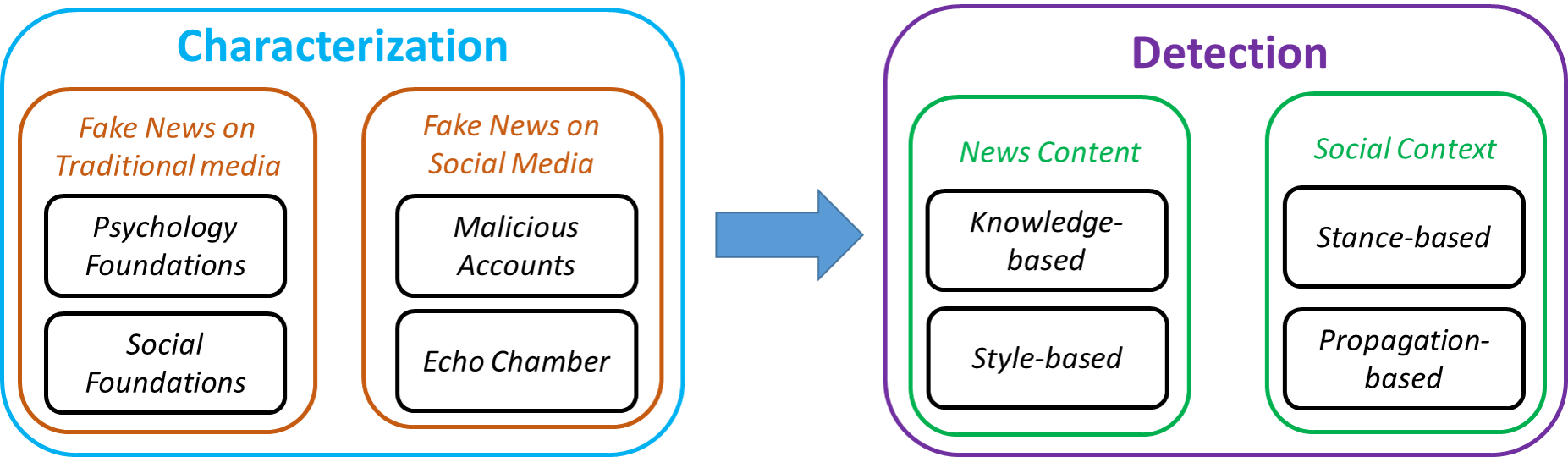}
\caption{Fake news on social media: from characterization to detection.}\label{fig:framework}
\end{figure*}

To facilitate  research in fake news detection on social media, in this survey we will review two aspects of the fake news detection problem: 
\textit{characterization} and \textit{detection}. As shown in Figure~\ref{fig:framework}, we will first describe the background of the fake news detection problem using theories and properties from psychology and social studies; then we present the detection approaches. Our major contributions of this survey are summarized as follows:
\begin{itemize}
\item We discuss the narrow and broad definitions of fake news that cover most existing definitions in the literature and further present the unique characteristics of fake news on social media and its implications compared with the traditional media;
\item We give an overview of existing fake news detection methods with a principled way to group representative methods into different categories; and
\item We discuss several open issues and provide future directions of fake news detection in social media.
\end{itemize}

The remainder of this survey is organized as follows. In Section~\ref{FN_Chara}, we present the definition of fake news and characterize it by comparing different theories and properties in both traditional and social media. In Section~\ref{FN_dete}, we continue to formally define the fake news detection problem  and summarize the methods to detect fake news. In Section~\ref{sec:eval}, we discuss the datasets and evaluation metrics used by existing methods. We briefly introduce  areas related to fake news detection on social media in Section~\ref{sec:related}. Finally, we discuss the open issues and future directions in Section~\ref{sec:future} and conclude this survey in Section~\ref{sec:conclusion}.

\section{Fake news characterization} \label{FN_Chara}
In this section, we introduce the basic social and psychological theories related to fake news and discuss more advanced patterns introduced by social media. Specifically, we first discuss various definitions of fake news and differentiate related concepts that are usually misunderstood as fake news. We then describe different aspects of fake news on traditional media and the new patterns found on social media.

\subsection{Definitions of Fake News}
Fake news has existed for a very long time, nearly the same amount of time as news began to circulate widely after the printing press was invented in 1439\footnote{http://www.politico.com/magazine/story/2016/12/fake-news-history-long-violent-214535}. However, there is no agreed definition of the term ``fake news''.  Therefore, we first discuss and compare some widely used definitions of fake news in the existing literature, and provide our definition of fake news that will be used for the remainder of this survey.

A narrow definition of fake news is  news articles that are intentionally and verifiably false and could mislead readers~\cite{allcott2017social}. There are two key features of this definition: \textit{authenticity} and \textit{intent}. First, fake news includes false information that can be verified as such. Second, fake news is created with dishonest intention to mislead consumers. This definition has been widely adopted in recent studies~\cite{mustafaraj2017fake,conroy2015automatic,potthast2017stylometric,klein2017fake}. Broader definitions of fake news  focus on the either authenticity or intent of the news content. Some papers regard satire news as fake news since the contents are false even though satire is often entertainment-oriented and reveals its own deceptiveness to the consumers~\cite{rubin2016fake,balmas2014fake,jin2016news,brewer2013impact}. Other literature directly treats deceptive news as fake news~\cite{rubin2015deception}, which includes serious fabrications, hoaxes, and satires.

In this article, we use the narrow definition of fake news. Formally, we state this definition as  follows,
\\\\
\textsc{Definition 1}~~\textsc{(Fake News)}
\textit{Fake news is a news article that is intentionally and verifiably false.}
\\\\
The reasons for choosing this narrow definition are three-folds. First, the underlying intent of fake news provides both theoretical and practical value that enables a  deeper understanding and analysis of this topic. Second, any techniques for truth verification that apply to the narrow conception of fake news can also be applied to  under the broader definition. Third, this definition is able to eliminate the ambiguities between fake news and related concepts that are not considered in this article. The following concepts are not fake news according to our definition: (1) satire news with proper context, which has no intent to mislead or deceive consumers and is unlikely to be mis-perceived as factual; (2) rumors that did not originate from news events; (3) conspiracy theories, which are difficult verify as true or false; (4) misinformation that is created unintentionally; and (5) hoaxes that are only motivated by fun or to scam targeted individuals. 



\subsection{Fake News on Traditional News Media}
Fake news itself is not a new problem. The media ecology of fake news has been changing over time from  newsprint to radio/television and, recently, online news and social media. We denote ``traditional fake news'' as the fake news problem before social media had important effects on its production and dissemination. Next, we will describe several psychological and social science foundations that describe the impact of fake news at both the individual and social information ecosystem levels.
\ \\
\ \\
\textbf{Psychological Foundations of Fake News.} Humans are naturally not very good at differentiating between real and fake news. There are several psychological and cognitive theories that can explain this phenomenon and the influential power of fake news. Traditional fake news mainly targets consumers by exploiting their \textit{individual vulnerabilities}. There are two major factors which make consumers  naturally vulnerable to fake news: (i) \textit{Na\"{i}ve Realism}: consumers tend to believe that their perceptions of reality are the only accurate views, while others who disagree are regarded as uninformed, irrational, or biased~\cite{ward1997naive}; and (ii) \textit{Confirmation Bias}: consumers prefer to receive information that confirms their existing views~\cite{nickerson1998confirmation}.  Due to these cognitive biases inherent in human nature, fake news can often be perceived as real by consumers. Moreover, once the misperception is formed, it is very hard to correct it. Psychology studies shows that correction of false information (e.g., fake news) by the presentation  of true, factual information is not only unhelpful to reduce misperceptions, but sometimes may even increase the misperceptions, especially among  ideological groups~\cite{nyhan2010corrections}.
\ \\

\textbf{Social  Foundations  of the Fake News Ecosystem.} Considering the entire news consumption ecosystem, we can also describe some of the social dynamics that contribute to the proliferation of fake news. Prospect theory describes decision making as a process by which people make choices based on the relative gains and losses as compared to their current state~\cite{kahneman1979, tversky1992advances}. This desire for maximizing the reward of a decision applies to social gains as well, for instance, continued acceptance by others in a user's immediate social network. As described by social identity theory~\cite{tajfel1979integrative, tajfel2004social} and normative influence theory~\cite{asch1951effects, kapferer2017rumors}, this preference for social acceptance and affirmation is essential to a person's identity and self-esteem, making users likely to choose ``socially safe'' options when consuming and disseminating news information, following the norms established in the community even if the news being shared is fake news. 

This rational theory of fake news interactions can be modeled from an economic game theoretical perspective~\cite{gentzkow2014media} by formulating the news generation and consumption cycle as a two-player strategy game.  For explaining fake news, we assume there are two kinds of key players in the information ecosystem:  \textit{publisher} and \textit{consumer}. The process of news publishing is modeled as a mapping from original signal $s$ to resultant news report $a$ with an effect of distortion bias $b$, i.e., $s \xrightarrow{b} a$, where $b=[-1,0,1]$ indicates $[left, no, right]$ biases take effects on news publishing process. Intuitively, this is capturing the degree to which a news article may be biased or distorted to produce fake news. The utility for the  publisher stems from two perspectives: (i) \textit{short-term utility:} the incentive to maximize profit, which is positively correlated with the number of consumers reached; (ii) \textit{long-term utility:} their reputation in terms of news authenticity.  Utility of consumers consists of two parts: (i) \textit{information utility}: obtaining true and unbiased information (usually extra investment cost needed); (ii) \textit{psychology utility}: receiving news that satisfies their prior opinions and social needs, e.g., confirmation bias and prospect theory. Both publisher and consumer try to maximize their overall utilities in this strategy game of the news consumption process. We can capture the fact that fake news happens when the \textit{short-term utility} dominates a publisher's overall utility and \textit{psychology utility} dominates the consumer's overall utility, and an equilibrium is maintained. This explains the social dynamics that lead to an information ecosystem where fake news can thrive. 


\subsection{Fake News on Social Media}\label{bots}
In this subsection, we will discuss some unique characteristics of fake news on social media. Specifically, we will highlight the key features of fake news that are enabled by social media. Note that the aforementioned characteristics of traditional fake news are also applicable to social media.

\ \\
\textbf{Malicious Accounts on Social Media for Propaganda.} While many users on social media are legitimate, social media users may also be malicious, and in some cases are not even real humans. The low cost of creating social media accounts also encourages malicious user accounts, such as social bots, cyborg users, and trolls. A  social bot refers to a social media account that is controlled by a computer algorithm to automatically produce content and interact with humans (or other bot users) on social media~\cite{ferrara2016rise}. Social bots can become malicious entities designed specifically with the purpose to do harm, such as manipulating and spreading fake news on social media. Studies shows that social bots  distorted the 2016 U.S. presidential election online discussions on a large scale~\cite{bessi2016social}, and that around 19 million bot accounts tweeted in support of either Trump or Clinton in the week leading up to election day\footnote{http://comprop.oii.ox.ac.uk/2016/11/18/resource-for-understanding-political-bots/}. Trolls,  real human users who aim to disrupt online communities and provoke consumers into an emotional response,  are also playing an important role in spreading fake news on social media. For example, evidence suggests that there were 1,000 paid Russian trolls spreading fake news on Hillary Clinton\footnote{http://www.huffingtonpost.com/entry/russian-trolls-fake-news\_us\_58dde6bae4b08194e3b8d5c4}.  Trolling behaviors are highly affected by people's mood and the context of online discussions, which enables the easy dissemination of fake news among otherwise ``normal'' online communities~\cite{Cheng2017}. The effect of trolling is to trigger people's inner negative emotions, such as anger and fear, resulting in  doubt,  distrust, and irrational behavior. Finally, cyborg users can spread fake news in a way that blends automated activities with human input. Usually cyborg accounts are registered by human as a camouflage and set automated programs to perform activities in social media. The easy switch of functionalities between human and bot offers cyborg users unique opportunities to spread fake news~\cite{chu2012detecting}. In a nutshell, these highly active and partisan malicious accounts on social media become the powerful sources and proliferation of fake news.
\ \\
\ \\
\textbf{Echo Chamber Effect.} Social media provides a new paradigm of information creation and consumption for users. The information seeking and consumption process are changing from a mediated form (e.g., by journalists) to a more disinter-mediated way~\cite{del2016echo}. Consumers are selectively exposed to certain kinds of news because of the way  news feed appear on their homepage in social media, amplifying the psychological challenges to dispelling fake news identified above. For example, users on Facebook always follow like-minded people and thus receive news that promote their favored existing narratives~\cite{quattrociocchi2016echo}. Therefore, users on social media tend to form groups containing like-minded people where they then polarize their opinions, resulting in an \textit{echo chamber} effect. The echo chamber effect facilitates the process by which people consume and believe fake news due to the following psychological factors~\cite{paulrussian}: (1) \textit{social credibility}, which means people are more likely to perceive a source as credible if others perceive the source  is credible, especially when there is not enough information available to access the truthfulness of the source; and (2) \textit{frequency heuristic}, which means that consumers may naturally favor information they hear frequently, even if it is fake news. Studies have shown that increased exposure to an idea is enough to generate a positive opinion of it~\cite{zajonc1968attitudinal, zajonc2001mere}, and in echo chambers, users continue to share and consume the same information.  As a result, this echo chamber effect  creates segmented, homogeneous communities with a very limited information ecosystem. Research shows that the homogeneous communities become the primary driver of information diffusion that further strengthens polarization~\cite{del2016spreading}.

\section{Fake News Detection} \label{FN_dete}

In the previous section, we introduced the conceptual characterization of traditional fake news and fake news in social media. Based on this characterization, we further explore the problem definition and  proposed approaches for fake news detection.

\subsection{Problem Definition}
In this subsection, we present the details of mathematical formulation of fake news detection on social media. Specifically, we will introduce the definition of key components of fake news and then present the formal definition of fake news detection. The basic notations are defined  below,

\begin{itemize}
\item Let $a$ refer to a \textit{News Article}. It consists of two major components: \textit{Publisher} and \textit{Content}. Publisher $\vec{p_a}$ includes a set of profile features to describe the original author, such as name, domain, age, among other attributes. Content $\vec{c_a}$  consists of a set of attributes that represent the news article and includes headline, text, image, etc.

\item We also define \textit{Social News Engagements} as a set of tuples $\mathcal{E}=\{e_{it}\}$ to represent the process of how news spread over time among $n$ users $\mathcal{U}=\{u_1,u_2,...,u_n\}$ and their corresponding posts $\mathcal{P}=\{p_1,p_2,...,p_n\}$ on social media regarding news article $a$. Each engagement $e_{it}=\{u_i,p_i,t\} $ represents that a user $u_i$ spreads news article $a$ using $p_i$ at time $t$. Note that we set $t=Null$ if the article $a$ does not have any engagement yet and thus $u_i$ represents the publisher.
\end{itemize}

\textsc{Definition 2}~~\textsc{(Fake News Detection)}
\textit{Given the social news engagements $\mathcal{E}$ among $n$ users for news article $a$, the task of fake news detection is to predict whether the news article $a$ is a fake news piece or not, i.e., $\mathcal{F}:\mathcal{E} \rightarrow \{0,1\}$ such that,} \\
\begin{equation}
\mathcal{F}(a) = \begin{cases} 1, & \mbox{if }  a \text{~is a piece of fake news,}\\ 0, & \mbox{otherwise.} \end{cases}
\end{equation}
\textit{where $\mathcal{F}$ is the prediction function we want to learn.}

Note that we define fake news detection as a binary classification problem for the following reason: fake news is essentially a \textit{distortion bias} on  information manipulated by the publisher. According to previous research about media bias theory~\cite{gentzkow2014media}, distortion bias is usually modeled as a binary classification problem.

Next, we propose a general data mining framework for fake news detection which includes two phases: (i) feature extraction and (ii) model construction. The feature extraction phase aims to represent news content and related auxiliary information in a formal mathematical structure, and model construction phase further builds machine learning models to better differentiate fake news and real news based on the feature representations.

\subsection{Feature Extraction}
Fake news detection on traditional news media mainly relies on news content, while in social media, extra social context auxiliary information can be used to as additional information to help detect fake news. Thus, we will present the details of how to extract and represent useful features from \textit{news content} and \textit{social context}.

\subsubsection{News Content Features}
News content features $\vec{c_a}$ describe the meta information related to a piece of news. A list of representative news content attributes are listed below:
\begin{itemize}
\item \textbf{Source:} Author or publisher of the news article
\item \textbf{Headline:}  Short title text that aims to catch the attention of readers and describes the main topic of the article 
\item \textbf{Body Text:} Main text that elaborates the details of the news story; there is usually a major claim that is specifically highlighted and that shapes the angle of the publisher
\item \textbf{Image/Video:} Part of the  body content of a news article that provides visual cues to frame the story 
\end{itemize}

Based on these raw content attributes, different kinds of feature representations can be built to extract discriminative characteristics of fake news. Typically, the news content we are looking at will mostly be \textit{linguistic-based} and \textit{visual-based}, described in more detail below. 

\ \\
\ \\
\textbf{Linguistic-based:} Since fake news pieces are intentionally created for financial or political gain rather than to report objective claims, they often contain opinionated and inflammatory language, crafted as ``clickbait'' (i.e., to entice users to click on the link to read the full article) or to incite confusion~\cite{chen2015misleading}. Thus, it is reasonable to exploit linguistic features that capture the different writing styles and sensational headlines to detect fake news. Linguistic-based features are extracted from the text content in terms of document organizations from different levels, such as characters, words, sentences, and documents.  In order to capture the different aspects of fake news and real news, existing work utilized both common linguistic features and domain-specific linguistic features. Common linguistic features are often used to represent documents for various tasks in natural language processing. Typical common linguistic features are: (i) \textit{lexical features}, including character-level and word-level features, such as total words, characters per word, frequency of large words, and unique words; (ii) \textit{syntactic features}, including sentence-level features, such as frequency of function words and phrases (i.e., ``n-grams'' and bag-of-words approaches~\cite{furnkranz1998study}) or punctuation and parts-of-speech (POS) tagging. Domain-specific linguistic features, which are specifically aligned to news domain, such as quoted words, external links, number of graphs, and the average length of graphs, etc~\cite{potthast2017stylometric}. Moreover, other features can be specifically designed to capture the deceptive cues in writing styles to differentiate fake news, such as lying-detection features~\cite{afroz2012detecting}.  
\ \\
\ \\
\textbf{Visual-based}: Visual cues have been shown to be an important manipulator for fake news propaganda\footnote{https://www.wired.com/2016/12/photos-fuel-spread-fake-news/}. As we have characterized, fake news exploits the individual vulnerabilities of people and thus often relies on sensational or even fake images to provoke anger or other emotional response of consumers. Visual-based features are extracted from visual elements (e.g. images and videos) to capture the different characteristics for fake news. Faking images were identified based on various user-level and tweet-level hand-crafted features using classification framework~\cite{gupta2013faking}. Recently, various \textit{visual} and \textit{statistical} features has been extracted for news verification~\cite{jin2017novel}. Visual features include clarity score, coherence score, similarity distribution histogram, diversity score, and clustering score. Statistical features include count, image ratio, multi-image ratio, hot image ratio, long image ratio, etc.

\subsubsection{Social Context Features}
In addition to features related directly to the content of the news articles, additional social context features can also be derived from the user-driven social engagements of news consumption on social media platform. Social engagements represent the news proliferation process over time, which provides useful auxiliary information to infer the veracity of  news articles. Note that  few papers exist in the literature that detect fake news using social context features. However, because we believe this is a critical aspect of successful fake news detection, we introduce a set of common features utilized in similar research areas, such as rumor veracity classification on social media. Generally, there are three major aspects of the social media context that we want to represent:  users, generated posts, and networks. Below, we investigate how we can extract and represent social context features from these three aspects to support fake news detection.
\ \\
\ \\
\textbf{User-based}: As we mentioned in Section~\ref{bots}, fake news pieces are likely to be created and spread by non-human accounts, such as social bots or cyborgs. Thus, capturing users' profiles and characteristics by user-based features can provide useful information for fake news detection. User-based features represent the characteristics of those users who have interactions with the news on social media. These features can be categorized across different levels: \textit{individual level} and \textit{group level}. Individual level  features are extracted to infer the credibility and reliability for each user using various aspects of user demographics, such as registration age, number of followers/followees, number of tweets the user has authored, etc~\cite{castillo2011information}. Group level user features capture overall characteristics of groups of users related to the news~\cite{yang2012automatic}. The assumption is that the spreaders of fake news and real news may form different communities with unique characteristics that can be depicted by group level features. Commonly used group level features come from aggregating (e.g., averaging and weighting) individual level features, such as `percentage of verified users' and `average number of followers'~\cite{ma2015detect,kwon2013prominent}.
\ \\
\ \\
\textbf{Post-based:} People express their emotions or opinions towards fake news through social media posts, such as skeptical opinions, sensational reactions, etc. Thus, it is reasonable to extract post-based features to help find potential fake news via reactions from the general public as expressed in posts. Post-based features focus on identifying useful information to infer the veracity of news from various aspects of relevant social media posts. These features can be categorized as \textit{post level}, \textit{group level}, and \textit{temporal level}. Post level features generate feature values for each post. The aforementioned linguistic-based features and some embedding approaches~\cite{ruchansky2017csi} for news content can also be applied for each post. Specifically, there are unique features for posts that represent the social response from general public, such as \textit{stance}, \textit{topic}, and \textit{credibility}. Stance features (or viewpoints) indicate the users' opinions towards the news, such as supporting, denying, etc~\cite{jin2016news}. Topic features can be extracted using topic models, such as latent Dirichlet allocation (LDA)~\cite{ma2015detect}. Credibility features for posts assess the degree of reliability~\cite{castillo2011information}. Group level features aim to aggregate the feature values for all relevant posts for specific news articles by using ``wisdom of crowds''. For example, the average credibility scores are used to evaluate the credibility of news~\cite{jin2016news}. A more comprehensive list of group-level post features can also be found in~\cite{castillo2011information}. Temporal level features consider the temporal variations of post level feature values~\cite{ma2015detect}. Unsupervised embedding methods, such as recurrent neural network (RNN), are utilized to capture the changes in posts over time~\cite{ruchansky2017csi,ma2016detecting}. Based on the shape of this time series for various metrics of relevant posts  (e.g,  number of posts), mathematical features can be computed, such as SpikeM parameters~\cite{kwon2013prominent}.
\ \\
\ \\
\textbf{Network-based}: Users form different networks on social media in terms of interests, topics, and relations. As mentioned before, fake news dissemination processes tend to form an echo chamber cycle, highlighting the value of  extracting network-based features to represent these types of network patterns for fake news detection. Network-based features are extracted via constructing specific networks among the users who published related social media posts. Different types of networks can be constructed. The \textit{stance network} can be built with nodes indicating all the tweets relevant to the news and the edge indicating the weights of similarity of stances~\cite{jin2016news,tacchini2017some}. Another type of network is the \textit{co-occurrence network}, which is built based on the user engagements by counting whether those users write posts relevant to the same news articles~\cite{ruchansky2017csi}. In addition, the \textit{friendship network} indicates the following/followee structure of users who post related tweets~\cite{kwon2013prominent}. An extension of this friendship network is  the \textit{diffusion network}, which tracks the trajectory of the spread of news~\cite{kwon2013prominent}, where nodes represent the users and edges represent the information diffusion paths among them. That is, a diffusion path between two users $u_i$ and  $u_j$ exists if and only if (1) $u_j$ follows $u_i$, and (2) $u_j$ posts about a given news only after $u_i$ does so. After these networks are properly built, existing network metrics can be applied as feature representations. For example, degree and clustering coefficient have been used to characterize the diffusion network~\cite{kwon2013prominent} and friendship network~\cite{kwon2013prominent}. Other approaches learn the latent node embedding features by using SVD~\cite{ruchansky2017csi} or network propagation algorithms~\cite{jin2016news}.

\subsection{Model Construction}
In the previous section, we introduced features extracted from different sources, i.e., news content and social context, for fake news detection. In this section, we discuss the details of the model construction process for several existing approaches. Specifically we categorize existing methods based on their main input sources as: \textit{News Content Models} and \textit{Social Context Models}.

\subsubsection{News Content Models}
In this subsection, we focus on news content models, which mainly rely on news content features and existing factual sources to classify fake news. Specifically, existing approaches can be categorized as  \textit{Knowledge-based} and \textit{Style-based}.
\ \\

\textbf{Knowledge-based:} Since fake news attempts to spread false claims in news content, the most straightforward means of detecting it  is to check the truthfulness of major claims in a news article to decide the news veracity. Knowledge-based approaches aim to use external sources to fact-check proposed claims in news content. The goal of fact-checking  is to assign a truth value to a claim in a particular context~\cite{vlachos2014fact}. Fact-checking has attracted increasing attention, and many efforts have been made to develop a feasible automated fact-checking system. Existing fact-checking approaches can be categorized as \textit{expert-oriented}, \textit{crowdsourcing-oriented}, and \textit{computational-oriented}.

\begin{itemize}
\item \textit{Expert-oriented} fact-checking heavily relies on human domain experts to investigate relevant data and documents to construct the verdicts of claim veracity, for example PolitiFact\footnote{http://www.politifact.com/}, Snopes\footnote{http://www.snopes.com/}, etc. However, expert-oriented fact-checking is an intellectually demanding and time-consuming process, which limits the potential for high efficiency and scalability.

\item \textit{Crowdsourcing-oriented} fact-checking exploits the ``wisdom of crowd" to enable normal people to annotate news content; these annotations are then aggregated to produce an overall assessment of the news veracity. For example, Fiskkit\footnote{http://fiskkit.com} allows users to discuss and annotate the accuracy of specific parts of a news article. As another example, an anti-fake news bot named ``For real'' is a public account in the instant communication mobile application LINE\footnote{https://grants.g0v.tw/projects/588fa7b382223f001e022944}, which allows people to report suspicious news content which is then further checked by editors.

\item \textit{Computational-oriented} fact-checking aims to provide an automatic scalable system to classify true and false claims.  Previous computational-oriented fact checking methods try to solve two majors issues: (i) identifying check-worthy claims and (ii) discriminating the veracity of fact claims. To identify check-worthy claims, factual claims in news content are extracted that convey key statements and viewpoints, facilitating the subsequent fact-checking process~\cite{hassan2015detecting}. Fact-checking for specific claims largely relies on \textit{external resources} to determine the truthfulness of a particular claim. Two typical external sources include the \textit{open web} and structured \textit{knowledge graph}. Open web sources are utilized as references that can be compared  with given claims in terms of both the consistency and frequency~\cite{banko2007open,magdy2010web}. Knowledge graphs are integrated from the linked open data as a structured network topology, such as DBpedia and Google Relation Extraction Corpus. Fact-checking using a knowledge graph aims to check whether the claims in news content can be inferred from existing facts in the knowledge graph~\cite{wu2014toward,ciampaglia2015computational,shi2016fact}.

\end{itemize}


\textbf{Style-based:} Fake news publishers often have malicious intent to spread distorted and misleading information and influence large communities of consumers, requiring particular writing styles necessary to appeal to and persuade a wide scope of consumers that is not seen in true news articles. Style-based approaches try to detect fake news by capturing the \textit{manipulators}  in the writing style of news content. There are mainly two typical categories of style-based methods: \textit{Deception-oriented} and \textit{Objectivity-oriented}.

\begin{itemize}
\item \textit{Deception-oriented} stylometric methods capture the deceptive statements or claims from news content. The motivation of deception detection originates from  forensic psychology (i.e., Undeutsch Hypothesis) ~\cite{undeutsch1967beurteilung} and various forensic tools including Criteria-based Content Analysis~\cite{vrij2005criteria} and Scientific-based Content Analysis~\cite{lesce1990scan} have been developed. More recently, advanced natural language processing models are applied to spot deception phases from the following perspectives: \textit{Deep syntax} and \textit{Rhetorical structure}. Deep syntax models have been implemented using probabilistic context frree grammers (PCFG), with which sentences can be transformed into rules that describe the syntax structure. Based on the PCFG, different rules can be developed for deception detection, such as unlexicalized/ lexicalized production rules and grandparent rules~\cite{feng2012syntactic}. Rhetorical structure theory can be utilized to capture the differences between deceptive and truthful sentences~\cite{rubin2015truth}. Deep network models, such as convolutional neural networks (CNN), have also been applied to classify fake news veracity~\cite{wang2017liar}.

\item \textit{Objectivity-oriented} approaches capture style signals that can indicate a decreased  objectivity of news content and thus the potential to mislead consumers, such as hyperpartisan styles and yellow-journalism. Hyperpartisan styles represent extreme behavior  in favor of a particular political party, which often correlates with a strong motivation to create fake news. Linguistic-based features can be applied to detect hyperpartisan articles~\cite{potthast2017stylometric}. Yellow-journalism represents those articles that do not contain well-researched news, but instead rely on eye-catching headlines (i.e., clickbait) with a propensity for exaggeration, sensationalization, scare-mongering, etc.  Often, news titles will summarize the major viewpoints of the article that the author wants to convey, and thus misleading and deceptive clickbait titles can serve as a good indicator for recognizing fake news articles~\cite{chen2015misleading}.

\end{itemize}

\subsubsection{Social Context Models}
The nature of social media provides researchers with additional resources to supplement and enhance News Content Models. Social context models include relevant user social engagements in the analysis, capturing this auxiliary information from a variety of perspectives. We can classify existing approaches for social context modeling  into two categories: \textit{Stance-based} and \textit{Propagation-based}. Note that very few existing fake news detection approaches have utilized social context models. Thus, we also introduce similar methods for rumor detection using social media, which  have potential application for fake news detection.
\ \\

\textbf{Stance-based:} Stance-based approaches utilize users' viewpoints from relevant post contents to infer the veracity of original news articles. The stance  of users' posts can be represented either \textit{explicitly} or \textit{implicitly}. Explicit stances are direct expressions of emotion or opinion, such as the ``thumbs up'' and ``thumbs down'' reactions expressed in Facebook. Implicit stances can be automatically extracted from social media posts. Stance detection is the task of automatically determining from a post whether the user is in favor of, neutral toward, or against some target entity, event, or idea~\cite{mohammad2017stance}. Previous stance classification methods mainly rely on hand-crafted linguistic or embedding features on individual posts to predict stances~\cite{mohammad2017stance,qazvinian2011rumor}. Topic model methods, such as latent dirichlet allocation (LDA) can  be applied to learn latent stance from topics~\cite{jin2016news}. Using these methods, we can  infer the news veracity based on the stance values of relevant posts.  Tacchini~\textit{et al.} proposed to construct a bipartite network of user and Facebook posts using the ``like'' stance information~\cite{tacchini2017some}; based on this network, a semi-supervised probabilistic model was used to predict the likelihood of Facebook posts being hoaxes. Jin~\textit{et al.} explored topic models to learn latent viewpoint values and further exploited these viewpoints to learn the credibility of relevant posts and  news content~\cite{jin2016news}.
\ \\

\textbf{Propagation-based:} Propagation-based approaches for fake news detection reason about the \textit{interrelations} of relevant social media posts to predict news credibility. The basic assumption is that the credibility of a news event is highly related to the credibilities of relevant social media posts. Both \textit{homogeneous} and \textit{heterogeneous} credibility networks can be built for propagation process. Homogeneous credibility networks consist of a single type of entities, such as post or event~\cite{jin2016news}. Heterogeneous credibility networks involve different types of entities, such as posts, sub-events, and events~\cite{jin2014news,gupta2012evaluating}. Gupta~\textit{et al.} proposed a PageRank-like credibility propagation algorithm by encoding users' credibilities and tweets' implications on a three layer user-tweet-event heterogeneous information network. Jin~\textit{et al.} proposed to include news aspects (i.e., latent sub-events), build a three-layer hierarchical network, and utilize a graph optimization framework to infer event credibilities. Recently, the conflicting viewpoint relationships are included to build a homogeneous credibility network among tweets and guide the process to evaluate their credibilities~\cite{jin2016news}.

\section{Assessing Detection Efficacy} \label{sec:eval}
In this section, we discuss how to assess the performance of algorithms for fake news detection. We focus on the available datasets and evaluation metrics for this task.

\subsection{Datasets}
Online news can be collected from different sources, such as news agency homepages, search engines, and social media websites. However, manually determining the veracity of news is a challenging task, usually requiring  annotators with domain expertise who performs careful analysis of claims and additional evidence, context, and reports from  authoritative sources. Generally, news data with annotations can be gathered in the following ways: \textit{Expert journalists}, \textit{Fact-checking websites}, \textit{Industry detectors}, and \textit{Crowdsourced workers}. However, there are no agreed upon benchmark datasets for the fake news detection problem. Some publicly available datasets are listed below:

\begin{itemize}
\item \textit{BuzzFeedNews}\footnote{https://github.com/BuzzFeedNews/2016-10-facebook-fact-check/tree/master/data}: This dataset comprises a complete sample of news published in Facebook from 9 news agencies over a week close to the 2016 U.S. election from September 19 to 23 and September 26 and 27. Every post and the linked article were fact-checked claim-by-claim by 5 BuzzFeed journalists. This dataset is further enriched in~\cite{potthast2017stylometric} by adding the linked articles, attached media, and relevant metadata. It contains 1,627 articles--826 mainstream, 356 left-wing, and 545 right-wing articles.

\item \textit{LIAR}\footnote{https://www.cs.ucsb.edu/~william/data/liar\_dataset.zip}: This dataset is collected from  fact-checking website  PolitiFact through its API~\cite{wang2017liar}. It includes  12,836 human-labeled short statements, which are sampled from various contexts, such as news releases, TV or radio interviews, campaign speeches, etc. The labels for  news truthfulness are fine-grained multiple classes: pants-fire, false, barely-true, half-true, mostly true, and true.

\item \textit{BS Detector}\footnote{https://www.kaggle.com/mrisdal/fake-news}: This dataset is collected from a browser extension called BS detector developed for checking news veracity\footnote{https://github.com/bs-detector/bs-detector}. It searches all links on a given webpage for references to unreliable sources by checking against a manually complied list of domains. The labels are the outputs of BS detector, rather than human annotators.

\item \textit{CREDBANK}\footnote{http://compsocial.github.io/CREDBANK-data/}: This is a large scale crowdsourced dataset of approximately 60 million tweets that cover 96 days starting from October 2015. All the tweets are broken down to be related to over 1,000 news events, with each event assessed for credibilities by 30 annotators from Amazon Mechanical Turk~\cite{mitra2015credbank}.

\end{itemize}

In Table~\ref{tab:data}, we compare these public fake news detection datasets, highlighting the features that can be extracted from each dataset. We can see that no existing public dataset can provide all possible features of interest. In addition, these datasets also have specific limitation that make them challenging to use for fake news detection. BuzzFeedNews only contains headlines and text for each news piece and  covers news articles from very few news agencies.  LIAR includes mostly short statements, rather than the entire news content. Further, these statements are collected from various speakers, rather than news publishers, and may include some claims that are not fake news. BS Detector data is collected and annotated by using a developed news veracity checking tool. As the labels have not been properly validated by human experts, any model trained on this data is really learning the parameters of BS Detector, rather than expert-annotated ground truth fake news. Finally, CREDBANK was originally collected for tweet credibility assessment, so the  tweets in this dataset are not really the social engagements for specific news articles.

To address the disadvantages of existing fake news detection datasets, we have an ongoing project to develop a usable dataset for fake news detection on social media. This dataset, called $FakeNewsNet$\footnote{https://github.com/KaiDMML/FakeNewsNet},  includes all mentioned news content and social context features with reliable ground truth fake news labels.

\begin{table*} [tbp!]
\centering \caption{Comparison of Fake News Detection Datasets.}
\vskip -1em
\begin{tabular}{l|ccccc}
\toprule
 \multirow{2}{*}{\diagbox{Dataset}{Features}} &\multicolumn{2}{c}{\textbf{News Content}} &\multicolumn{3}{c}{\textbf{Social Context}} \\
\cline{2-6}
& \textbf{Linguistic} & \textbf{Visual} & \textbf{User} & \textbf{Post} & \textbf{Network}  \\
\midrule
 \textbf{BuzzFeedNews} & \checkmark & & & & \\
\midrule
\textbf{LIAR} & \checkmark & & & & \\
\midrule
\textbf{BS Detector} & \checkmark & & & & \\
\midrule
\textbf{CREDBANK} & \checkmark & &\checkmark &\checkmark & \checkmark\\
\bottomrule
\end{tabular} \label{tab:data}
\vskip -1em
\end{table*}

\subsection{Evaluation Metrics}
To evaluate the performance of algorithms for fake news detection problem, various evaluation metrics have been used. In this subsection, we review the most widely used metrics for fake news detection. Most existing approaches consider the fake news problem as a classification problem that predicts whether a news article is fake or not:

\begin{itemize}
\vspace{-1mm}
\setlength{\itemsep}{0pt}
\item True Positive (\textbf{TP}): when predicted fake news pieces are actually annotated as fake news;
\item True Negative (\textbf{TN}): when predicted true news pieces are actually annotated as true news;
\item False Negative (\textbf{FN}): when predicted true news pieces are actually annotated as fake news;
\item False Positive (\textbf{FP}): when predicted fake news pieces are actually annotated as true news.
\end{itemize}
By formulating this as a classification problem, we can define following metrics,
\begin{eqnarray}
Precision  &=& \frac{|TP|}{|TP|+|FP|}\\
Recall  &=& \frac{|TP|}{|TP|+|FN|}\\
F1  &=&  2\cdot \frac{Precision \cdot Recall}{Precision+Recall}\\
Accuracy &=& \frac{|TP|+|TN|}{|TP|+|TN|+|FP|+|FN|}
\end{eqnarray}
These metrics are commonly used in the machine learning community and enable us to evaluate the performance of a classifier from different perspectives. Specifically, accuracy measures the similarity between predicted fake news and real fake news.  Precision measures the fraction of all detected fake news that are annotated as fake news, addressing the important problem of identifying which news is fake. However, because fake news datasets are often skewed, a high precision can be easily achieved by making fewer positive predictions. Thus, recall is used to measure the sensitivity, or the fraction of annotated fake news articles that are predicted to be fake news.  F1 is used to combine precision and recall, which can provide an overall prediction performance for fake news detection. Note that for $Precision, Recall$, $F_1$, and $Accuracy$, the higher the value, the better the performance.

The \textit{Receiver Operating Characteristics} (ROC) curve provides a way of comparing the performance of classifiers by looking at the trade-off in the \textit{False Positive Rate} (FPR) and the \textit{True Positive Rate} (TPR). To draw the ROC curve, we plot the FPR on the $x$ axis and and TPR along the $y$ axis. The ROC curve compares the performance of different classifiers by changing class distributions via a threshold. TPR and FPR are defined as follows (note that TPR is the same as recall defined above):
\begin{eqnarray}
TPR  &=& \frac{|TP|}{|TP|+|FN|}\\
FPR  &=& \frac{|FP|}{|FP|+|TN|}
\end{eqnarray}
Based on the ROC curve, we can compute the Area Under the Curve (AUC) value, which  measures the overall performance of how likely the classifier is to rank the fake news higher than any true news. Based on~\cite{hand2001simple}, AUC is defined as below.
\begin{equation}
AUC = \frac{\sum(n_0+n_1+1-r_i)-n_0(n_0+1)/2}{n_0n_1}
\end{equation}
where $r_i$ is the rank of $i_{th}$ fake news piece and $n_0$ ($n_1$) is the number of fake (true) news pieces.
It is worth mentioning that AUC is more statistically consistent and more discriminating than accuracy~\cite{ling2003auc}, and it is usually applied in an imbalanced classification problem, such as fake news classification, where the number of ground truth fake news articles and and true news articles have a very imbalanced distribution.

\section{Related Areas} \label{sec:related}
In this section, we further discuss areas that are related to the problem of fake news detection. We aim to point out the differences between these areas and fake news detection by briefly explaining the task goals and highlighting some popular methods.

\subsection{Rumor Classification}
A rumor can usually be defined as ``a piece of circulating information whose veracity status is yet to be verified at the time of spreading''~\cite{zubiaga2017detection}. The function of a rumor is to make sense of an \textit{ambiguous} situation, and the truthfulness value could be \textit{true}, \textit{false} or \textit{unverified}. Previous approaches for rumor analysis focus on four subtasks: rumor detection, rumor tracking, stance classification, and veracity classification~\cite{zubiaga2017detection}. Specifically, rumor detection aims to classify a piece of information as rumor or non-rumor~\cite{wu2017gleaning,sampson2016leveraging}; rumor tracking aims to collect and filter posts  discussing specific rumors; rumor stance classification determines how each relevant post is oriented with respect to the rumor's veracity; veracity classification attempts to predict the actual truth value of the rumor. The most related task to fake news detection is the rumor veracity classification. Rumor veracity classification relies heavily on the other subtasks, requiring the stances or opinions can be extracted from relevant posts.  These posts are considered as important sensors for determining the veracity of the rumor. Different from rumors, which may include long-term rumors, such as conspiracy theories, as well as short-term emerging rumors, fake news refers to  information related specifically to public news events that can be verified as false.

\subsection{Truth Discovery}
Truth discovery is the problem of detecting true facts from multiple conflicting sources~\cite{li2016survey}. Truth discovery methods do not explore the fact claims directly, but rely on a collection of contradicting sources that record the properties of objects to determine the truth value. Truth discovery aims to determine the \textit{source credibility} and \textit{object truthfulness} at the same time. The fake news detection problem can benefit from various aspects of  truth discovery approaches under different scenarios. First, the credibility of different news outlets can be modeled to infer the truthfulness of reported news. Second,  relevant social media posts can also be modeled as social response sources to better determine the truthfulness of claims~\cite{mukherjee2015leveraging,weikum2017computers}. However, there are some other issues that must be considered to apply truth discovery to fake news detection in social media scenarios. First, most existing truth discovery methods focus on handling \textit{structured} input in the form of Subject-Predicate-Object (SPO) tuples, while social media data is highly unstructured and noisy. Second, truth discovery methods can not be well applied when a fake news article is newly launched and published by only a few news outlets because at that point there is not enough social media posts relevant to it to serve as additional sources.

\subsection{Clickbait Detection}
Clickbait is a term commonly used to describe eye-catching and teaser headlines in online media. Clickbait headlines create a so-called ``curiosity gap'', increasing the likelihood that reader will click the target link to satisfy their curiosity. Existing clickbait detection approaches utilize various linguistic features extracted from teaser messages, linked webpages, and tweet meta information~\cite{chakraborty2016stop,blom2015click,potthast2016clickbait}. Different types of clickbait are categorized, and some of them are highly correlated with non-factual claims~\cite{biyani20168}. The underlying motivation of clickbait is usually for click-through rates and the resultant advertising revenue. Thus, the body text of clickbait articles are often informally organized and poorly reasoned. This discrepancy has been used by researchers to identify the inconsistency between headlines and news contents in an attempt to detect fake news articles\footnote{http://www.fakenewschallenge.org/}. Even though not all fake news may include clickbait headlines,  specific clickbait headlines could serve as an important indicator, and various features can be utilized to help  detect fake news.

\subsection{Spammer and Bot Detection}
Spammer detection on social media, which aims to capture malicious users that coordinate among themselves to launch various attacks, such as spreading ads, disseminating pornography, delivering viruses, and phishing~\cite{lee2010uncovering}, has recently attracted wide attention. Existing approaches for social spammer detection mainly rely on extracting features from user activities and social network information~\cite{hu2013social,wu2017adaptive,hu2014social,hu2014online}. In addition, the rise of social bots has also increased the circulation of false information as they automatically retweet posts without verifying the facts~\cite{ferrara2016rise}. The major challenge brought by social bots is that they can give a false impression that information is highly popular and endorsed by many people, which enables the echo chamber effect for the propagation of fake news. Previous approaches for bot detection are based on social network information, crowdsourcing, and discriminative features~\cite{ferrara2016rise,morstatter2016new,morstatter2016can}. Thus, both spammer and social bots could provide insights about target specific malicious social media accounts that can be used for fake news detection.

\section{Open Issues and Future Research} \label{sec:future}

\begin{figure*}[tb!]
\centering
\includegraphics[scale=0.45]{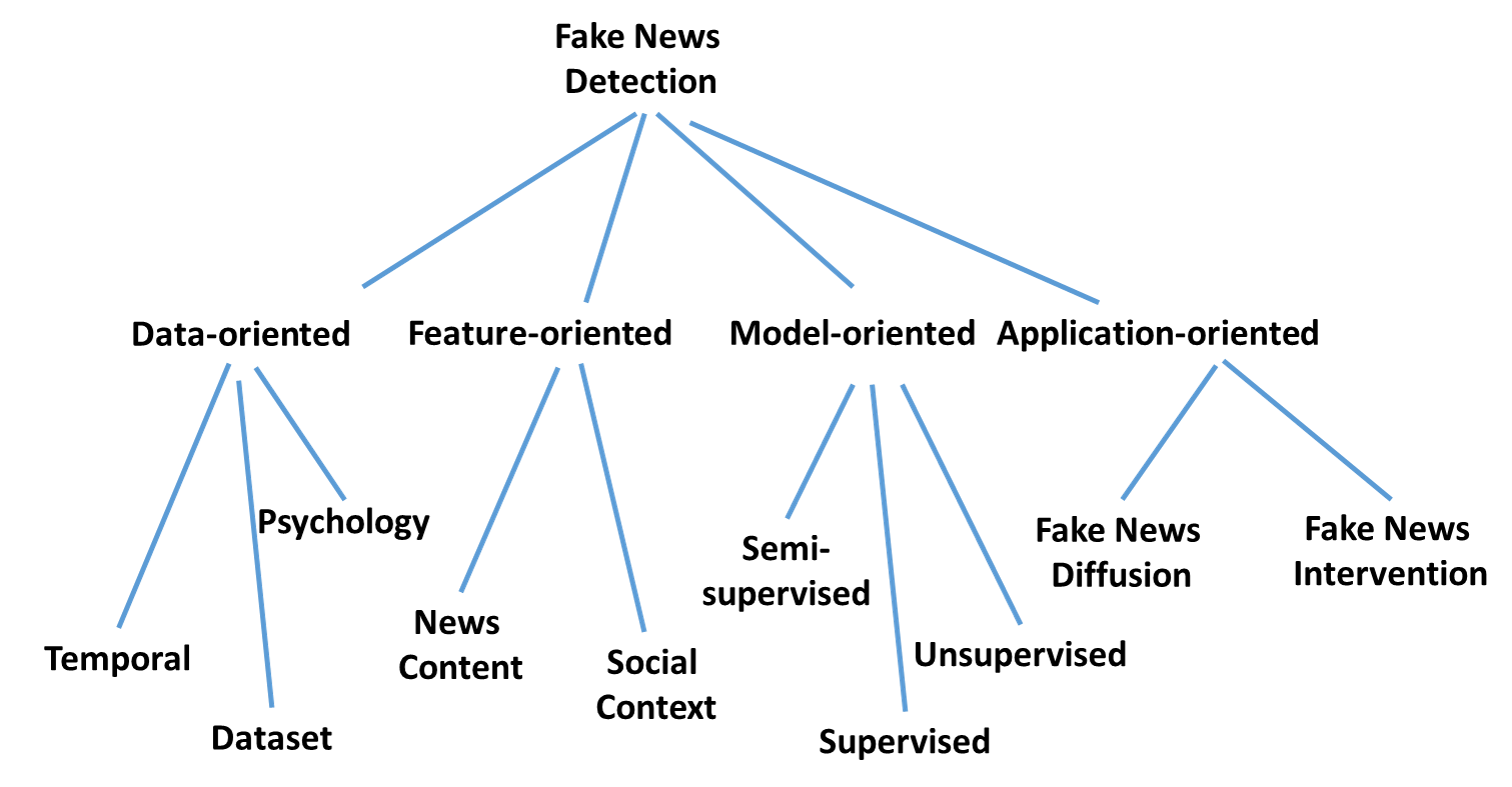}
\caption{Future directions and open issues for fake news detection on social media.}\label{fig:future}
\end{figure*}

In this section, we present some open issues in fake news detection and future research directions. Fake news detection on social media is a newly emerging research area, so we aim to point out promising research directions from a data mining perspective.
Specifically, as shown in Figure~\ref{fig:future}, we outline the research directions in four categories: \textit{Data-oriented}, \textit{Feature-oriented}, \textit{Model-oriented} and \textit{Application-oriented}.
\ \\
\ \\
\textbf{Data-oriented:} Data-oriented fake news research is focusing on different kinds of data characteristics,  such as : \textit{dataset}, \textit{temporal} and \textit{psychological}. From a dataset perspective, we demonstrated that there is no existing benchmark dataset that includes resources to extract all relevant features. A promising direction is to create a comprehensive and large-scale fake news benchmark dataset, which can be used by researchers to facilitate further research in this area. From a temporal perspective, fake news dissemination  on social media demonstrates unique temporal  patterns different from true news. Along this line, one interesting problem is to perform \textit{early fake news detection}, which aims to give early alerts of fake news during the dissemination process. For example, this approach could look at only social media posts within some time delay of the original post as sources for news verification~\cite{jin2016news}. Detecting fake news early can help prevent further propagation on social media. From a psychological perspective, different aspects of fake news have been qualitatively explored in the social psychology literature~\cite{ward1997naive,nickerson1998confirmation,nyhan2010corrections}, but quantitative studies to verify these psychological factors are rather limited. For example, the echo chamber effect plays an important role for fake news spreading in social media. Then how to capture echo chamber effects and how to utilize the pattern for fake news detection in social media could be an interesting investigation. Moreover, \textit{intention detection} from news data is promising but limited as most existing fake news research focus on detecting the authenticity but ignore the intent aspect of fake news. Intention detection is very challenging as the intention is often explicitly unavailable. Thus. it's worth to explore how to use data mining methods to validate and capture psychology intentions.
\ \\
\ \\
\textbf{Feature-oriented:} Feature-oriented fake news research aims to determine effective features for detecting fake news from multiple data sources. We have demonstrated that there are two major data sources: \textit{news content} and \textit{social context}. From a news content perspective, we introduced linguistic-based and visual-based techniques to extract features from text information. Note that linguistic-based features have been widely studied for general NLP tasks, such as text classification and clustering, and specific applications such as author identification~\cite{houvardas2006n} and deception detection~\cite{feng2012syntactic}, but the underlying characteristics of fake news have not been fully understood. Moreover, embedding techniques, such as word embedding and deep neural networks, are attracting much attention for textual feature extraction, and has the potential to learn better representations~\cite{wang2017liar,wang2016linked,wang2016paired}. In addition, \textit{visual features} extracted from images are also shown to be important indicators for fake news~\cite{jin2017novel}. However, very limited research has been done to exploit effective visual features, including traditional local and global features~\cite{ping2013review} and newly emerging deep network-based features~\cite{lecun2015deep,wang2017your,wang2017attributed}, for the fake news detection problem. Recently, it has been shown that advanced tools can manipulate video footage of public figures~\cite{thies2016face2face}, synthesize high quality videos~\cite{suwajanakorn2017synthesizing}, etc. Thus, it becomes much more challenging and important to differentiate real and fake visual content,  and more advanced visual-based features are needed for this research. From a social context perspective, we  introduced user-based, post-based, and network-based features. Existing user-based features mainly focus on general user profiles, rather than differentiating account types separately and extracting user-specific features. Post-based features can be represented using other techniques, such as convolutional neural networks (CNN)~\cite{ruchansky2017csi}, to better capture people's opinions and reactions toward fake news. Images in social media posts can also be utilized to better understand users' sentiments~\cite{wang2015unsupervised} toward news events. Network-based features are extracted to represent how different types of networks are constructed. It is important to extend this preliminary work to explore (i) how other networks can be constructed in terms of different aspects of relationships among relevant users and posts; and (ii) other advanced methods of network representations, such as network embedding~\cite{tang2015line,wang2017signed}.
\ \\
\ \\
\textbf{Model-oriented:} Model-oriented fake news research opens the door to building more effective and practical models for fake news detection. Most previously mentioned approaches focus on extracting various features, incorporating theses features into  supervised classification models,  such as na\"ive Bayes, decision tree, logistic regression, k nearest neighbor (KNN),  and support vector machines (SVM), and then selecting the classifier that performs the best~\cite{potthast2017stylometric,tacchini2017some,afroz2012detecting}. More research can be done to build more complex and effective models and to better utilize extracted features, such as \textit{aggregation methods}, \textit{probabilistic methods}, \textit{ensemble methods}, or \textit{projection methods}~\cite{shu2017user}. Specifically, we think there is some promising research  in the following directions. First, aggregation methods combine different feature representations into a weighted form and optimize the feature weights. Second, since fake news may commonly mix  true statements with false claims, it may make more sense to predict the likelihood of fake news instead of producing a binary value; probabilistic models  predict a probabilistic distribution of class labels (i.e., fake news versus true news) by assuming a generative model that pulls from the same distribution as the original feature space~\cite{garg2001understanding}.  Third, one of the major challenges for fake news detection is the fact that each feature, such as source credibility, news content style, or social response, has some limitations to directly predict fake news on its own. Ensemble methods  build a conjunction of several weak classifiers to learn a strong classifier that is more successful than any individual classifier alone; ensembles  have been widely applied to various applications in the machine learning literature~\cite{dietterich2000ensemble}. It may be beneficial to  build ensemble models as news content and social context features each have supplementary information that has the potential to boost fake news detection performance. Finally, fake news content or social context information may be noisy in the raw feature space; projection methods  refer to approaches that lean projection functions to map between original feature spaces (e.g., news content features and social context features) and the latent feature spaces that may be more useful for classification.

Moreover, most existing approaches are \textit{supervised}, which requires a pre-annotated fake news ground truth dataset to train a model. However, obtaining a reliable fake news dataset is very time and labor intensive, as the process often requires expert  annotators to perform careful analysis of claims and additional evidence, context, and reports from  authoritative sources. Thus, it is also important to consider scenarios where limited or no labeled fake news pieces are available in which \textit{semi-supervised} or \textit{unsupervised} models can be applied. While the models created by supervised classification methods may be more accurate given a well-curated ground truth dataset  for training, unsupervised models can be more practical because unlabeled datasets are easier to obtain.
\ \\
\ \\
\textbf{Application-oriented:} Application-oriented fake news research encompass research that goes into other areas beyond fake news detection. We propose two major directions along these lines: \textit{fake news diffusion} and \textit{fake news intervention}. Fake news diffusion characterizes the diffusion paths and patterns of fake news on social media sites. Some early research has shown that true information and misinformation follow different patterns when propagating  in online social networks~\cite{del2016spreading,menczer2016spread}. Similarly, the diffusion of fake news in social media demonstrates its own characteristics that need further investigation, such as \textit{social dimensions}, \textit{life cycle}, \textit{spreader identification}, etc. Social dimensions refer to the heterogeneity and weak dependency of social connections within different social communities. Users' perceptions of fake news pieces are highly affected by their like-minded friends in social media (i.e., echo chambers), while the \textit{degree} differs along  different social dimensions. Thus, it is worth  exploring why and how different social dimensions play a role in spreading fake news in terms of different topics, such as political, education, sports, etc. The fake news diffusion process also has different stages in terms of people's attentions and reactions as time goes by, resulting in a unique  life cycle. Research has shown that breaking news and in-depth news demonstrate different life cycles in social media~\cite{castillo2014characterizing}. Studying the life cycle of fake news will provide deeper understanding of how particular stories ``go viral'' from normal public discourse. Tracking the life cycle of fake news on social media requires recording essential trajectories of fake news diffusion in general~\cite{shao2016hoaxy}, as well as further investigations of the process for specific fake news pieces,  such as graph-based models and evolution-based models~\cite{guille2013information}. In addition, identifying key spreaders of fake news is crucial to mitigate the diffusion scope in social media. Note that key spreaders can be categorized in two ways, i.e., \textit{stance} and \textit{authenticity}. Along the stance dimensions, spreaders can either be (i) \textit{clarifiers}, who propose skeptical and opposing viewpoints towards fake news and try to clarify them; or (ii) \textit{persuaders}, who spread fake news with supporting opinions to persuade others to believe it. In this sense, it is important to explore how to detect clarifiers and persuaders and better use them to control the dissemination of fake news. From an authenticity perspective,  spreaders could be either \textit{human}, \textit{bot}, or \textit{cyborg}. Social bots have been used to  intentionally spread fake news in social media, which motivates further research to better characterize and detect malicious accounts designed for propaganda.

Finally, we also propose further research into fake news intervention, which aims to reduce the effects of fake news by \textit{proactive} intervention methods that minimize the spread scope or \textit{reactive} intervention methods after fake news goes viral. Proactive fake news intervention methods try to (i) remove malicious accounts that spread fake news or fake news itself to \textit{isolate} it from  future consumers; (ii) \textit{immunize} users with true news to change the belief of users that may already have been affected by fake news. There is recent research that attempts to use content-based immunization and network-based immunization methods in misinformation intervention~\cite{wu2016chapter,wu2016mining}. One approach uses a multivariate Hawkes process to model both true news and fake news and mitigate the spreading of fake news in real-time~\cite{farajtabar2017fake}. The aforementioned spreader detection techniques can also be applied to target certain users (e.g., persuaders) in social media to stop spreading fake news,  or other users (e.g. clarifiers) to maximize the spread of corresponding true news.

\section{Conclusion} \label{sec:conclusion}
With the increasing popularity of social media, more and more people consume news from social media instead of traditional news media. However, social media has also been used to spread fake news, which has strong negative impacts on individual users and broader society. In this article, we explored the fake news problem by reviewing existing literature in two phases: characterization and detection. In the characterization phase, we introduced the basic  concepts and principles of fake news in both traditional media and social media. In the detection phase, we reviewed existing fake news detection approaches from a data mining perspective, including feature extraction and model construction. We also further discussed the datasets, evaluation metrics, and promising future directions in fake news detection research and expand the field to other applications.

\section{Acknowledgments}
This material is based upon work supported by, or in part by, the ONR grant N00014-16-1-2257.

\balance
\bibliographystyle{plain}
\bibliography{abbrv}

\begin{thebibliography}{100}

\bibitem{afroz2012detecting}
Sadia Afroz, Michael Brennan, and Rachel Greenstadt.
\newblock Detecting hoaxes, frauds, and deception in writing style online.
\newblock In {\em ISSP'12}.

\bibitem{allcott2017social}
Hunt Allcott and Matthew Gentzkow.
\newblock Social media and fake news in the 2016 election.
\newblock Technical report, National Bureau of Economic Research, 2017.

\bibitem{asch1951effects}
Solomon~E Asch and H~Guetzkow.
\newblock Effects of group pressure upon the modification and distortion of
  judgments.
\newblock {\em Groups, leadership, and men}, pages 222--236, 1951.

\bibitem{balmas2014fake}
Meital Balmas.
\newblock When fake news becomes real: Combined exposure to multiple news
  sources and political attitudes of inefficacy, alienation, and cynicism.
\newblock {\em Communication Research}, 41(3):430--454, 2014.

\bibitem{banko2007open}
Michele Banko, Michael~J Cafarella, Stephen Soderland, Matthew Broadhead, and
  Oren Etzioni.
\newblock Open information extraction from the web.
\newblock In {\em IJCAI'07}.

\bibitem{bessi2016social}
Alessandro Bessi and Emilio Ferrara.
\newblock Social bots distort the 2016 us presidential election online
  discussion.
\newblock {\em First Monday}, 21(11), 2016.

\bibitem{biyani20168}
Prakhar Biyani, Kostas Tsioutsiouliklis, and John Blackmer.
\newblock " 8 amazing secrets for getting more clicks": Detecting clickbaits in
  news streams using article informality.
\newblock In {\em AAAI'16}.

\bibitem{blom2015click}
Jonas~Nygaard Blom and Kenneth~Reinecke Hansen.
\newblock Click bait: Forward-reference as lure in online news headlines.
\newblock {\em Journal of Pragmatics}, 76:87--100, 2015.

\bibitem{brewer2013impact}
Paul~R Brewer, Dannagal~Goldthwaite Young, and Michelle Morreale.
\newblock The impact of real news about “fake news”: Intertextual processes
  and political satire.
\newblock {\em International Journal of Public Opinion Research},
  25(3):323--343, 2013.

\bibitem{castillo2014characterizing}
Carlos Castillo, Mohammed El-Haddad, J{\"u}rgen Pfeffer, and Matt Stempeck.
\newblock Characterizing the life cycle of online news stories using social
  media reactions.
\newblock In {\em CSCW'14}.

\bibitem{castillo2011information}
Carlos Castillo, Marcelo Mendoza, and Barbara Poblete.
\newblock Information credibility on twitter.
\newblock In {\em WWW'11}.

\bibitem{chakraborty2016stop}
Abhijnan Chakraborty, Bhargavi Paranjape, Sourya Kakarla, and Niloy Ganguly.
\newblock Stop clickbait: Detecting and preventing clickbaits in online news
  media.
\newblock In {\em ASONAM'16}.

\bibitem{chen2015misleading}
Yimin Chen, Niall~J Conroy, and Victoria~L Rubin.
\newblock Misleading online content: Recognizing clickbait as false news.
\newblock In {\em Proceedings of the 2015 ACM on Workshop on Multimodal
  Deception Detection}, pages 15--19. ACM, 2015.

\bibitem{Cheng2017}
Justin Cheng, Michael Bernstein, Cristian Danescu-Niculescu-Mizil, and Jure
  Leskovec.
\newblock Anyone can become a troll: Causes of trolling behavior in online
  discussions.
\newblock In {\em CSCW '17}.

\bibitem{chu2012detecting}
Zi~Chu, Steven Gianvecchio, Haining Wang, and Sushil Jajodia.
\newblock Detecting automation of twitter accounts: Are you a human, bot, or
  cyborg?
\newblock {\em IEEE Transactions on Dependable and Secure Computing},
  9(6):811--824, 2012.

\bibitem{ciampaglia2015computational}
Giovanni~Luca Ciampaglia, Prashant Shiralkar, Luis~M Rocha, Johan Bollen,
  Filippo Menczer, and Alessandro Flammini.
\newblock Computational fact checking from knowledge networks.
\newblock {\em PloS one}, 10(6):e0128193, 2015.

\bibitem{conroy2015automatic}
Niall~J Conroy, Victoria~L Rubin, and Yimin Chen.
\newblock Automatic deception detection: Methods for finding fake news.
\newblock {\em Proceedings of the Association for Information Science and
  Technology}, 52(1):1--4, 2015.

\bibitem{del2016spreading}
Michela Del~Vicario, Alessandro Bessi, Fabiana Zollo, Fabio Petroni, Antonio
  Scala, Guido Caldarelli, H~Eugene Stanley, and Walter Quattrociocchi.
\newblock The spreading of misinformation online.
\newblock {\em Proceedings of the National Academy of Sciences},
  113(3):554--559, 2016.

\bibitem{del2016echo}
Michela Del~Vicario, Gianna Vivaldo, Alessandro Bessi, Fabiana Zollo, Antonio
  Scala, Guido Caldarelli, and Walter Quattrociocchi.
\newblock Echo chambers: Emotional contagion and group polarization on
  facebook.
\newblock {\em Scientific Reports}, 6, 2016.

\bibitem{dietterich2000ensemble}
Thomas~G Dietterich et~al.
\newblock Ensemble methods in machine learning.
\newblock {\em Multiple classifier systems}, 1857:1--15, 2000.

\bibitem{farajtabar2017fake}
Mehrdad Farajtabar, Jiachen Yang, Xiaojing Ye, Huan Xu, Rakshit Trivedi, Elias
  Khalil, Shuang Li, Le~Song, and Hongyuan Zha.
\newblock Fake news mitigation via point process based intervention.
\newblock {\em arXiv preprint arXiv:1703.07823}, 2017.

\bibitem{feng2012syntactic}
Song Feng, Ritwik Banerjee, and Yejin Choi.
\newblock Syntactic stylometry for deception detection.
\newblock In {\em ACL'12}.

\bibitem{ferrara2016rise}
Emilio Ferrara, Onur Varol, Clayton Davis, Filippo Menczer, and Alessandro
  Flammini.
\newblock The rise of social bots.
\newblock {\em Communications of the ACM}, 59(7):96--104, 2016.

\bibitem{furnkranz1998study}
Johannes F{\"u}rnkranz.
\newblock A study using n-gram features for text categorization.
\newblock {\em Austrian Research Institute for Artifical Intelligence},
  3(1998):1--10, 1998.

\bibitem{garg2001understanding}
Ashutosh Garg and Dan Roth.
\newblock Understanding probabilistic classifiers.
\newblock {\em ECML'01}.

\bibitem{gentzkow2014media}
Matthew Gentzkow, Jesse~M Shapiro, and Daniel~F Stone.
\newblock Media bias in the marketplace: Theory.
\newblock Technical report, National Bureau of Economic Research, 2014.

\bibitem{guille2013information}
Adrien Guille, Hakim Hacid, Cecile Favre, and Djamel~A Zighed.
\newblock Information diffusion in online social networks: A survey.
\newblock {\em ACM Sigmod Record}, 42(2):17--28, 2013.

\bibitem{gupta2013faking}
Aditi Gupta, Hemank Lamba, Ponnurangam Kumaraguru, and Anupam Joshi.
\newblock Faking sandy: characterizing and identifying fake images on twitter
  during hurricane sandy.
\newblock In {\em WWW'13}.

\bibitem{gupta2012evaluating}
Manish Gupta, Peixiang Zhao, and Jiawei Han.
\newblock Evaluating event credibility on twitter.
\newblock In {\em PSDM'12}.

\bibitem{hand2001simple}
David~J Hand and Robert~J Till.
\newblock A simple generalisation of the area under the roc curve for multiple
  class classification problems.
\newblock {\em Machine learning}, 2001.

\bibitem{hassan2015detecting}
Naeemul Hassan, Chengkai Li, and Mark Tremayne.
\newblock Detecting check-worthy factual claims in presidential debates.
\newblock In {\em CIKM'15}.

\bibitem{houvardas2006n}
John Houvardas and Efstathios Stamatatos.
\newblock N-gram feature selection for authorship identification.
\newblock {\em Artificial Intelligence: Methodology, Systems, and
  Applications}, pages 77--86, 2006.

\bibitem{hu2014social}
Xia Hu, Jiliang Tang, Huiji Gao, and Huan Liu.
\newblock Social spammer detection with sentiment information.
\newblock In {\em ICDM'14}.

\bibitem{hu2014online}
Xia Hu, Jiliang Tang, and Huan Liu.
\newblock Online social spammer detection.
\newblock In {\em AAAI'14}, pages 59--65, 2014.

\bibitem{hu2013social}
Xia Hu, Jiliang Tang, Yanchao Zhang, and Huan Liu.
\newblock Social spammer detection in microblogging.
\newblock In {\em IJCAI'13}.

\bibitem{jin2014news}
Zhiwei Jin, Juan Cao, Yu-Gang Jiang, and Yongdong Zhang.
\newblock News credibility evaluation on microblog with a hierarchical
  propagation model.
\newblock In {\em ICDM'14}.

\bibitem{jin2016news}
Zhiwei Jin, Juan Cao, Yongdong Zhang, and Jiebo Luo.
\newblock News verification by exploiting conflicting social viewpoints in
  microblogs.
\newblock In {\em AAAI'16}.

\bibitem{jin2017novel}
Zhiwei Jin, Juan Cao, Yongdong Zhang, Jianshe Zhou, and Qi~Tian.
\newblock Novel visual and statistical image features for microblogs news
  verification.
\newblock {\em IEEE Transactions on Multimedia}, 19(3):598--608, 2017.

\bibitem{kahneman1979}
Daniel Kahneman and Amos Tversky.
\newblock Prospect theory: An analysis of decision under risk.
\newblock {\em Econometrica: Journal of the econometric society}, pages
  263--291, 1979.

\bibitem{kapferer2017rumors}
Jean-Noel Kapferer.
\newblock {\em Rumors: Uses, Interpretation and Necessity}.
\newblock Routledge, 2017.

\bibitem{klein2017fake}
David~O Klein and Joshua~R Wueller.
\newblock Fake news: A legal perspective.
\newblock 2017.

\bibitem{kwon2013prominent}
Sejeong Kwon, Meeyoung Cha, Kyomin Jung, Wei Chen, and Yajun Wang.
\newblock Prominent features of rumor propagation in online social media.
\newblock In {\em ICDM'13}, pages 1103--1108. IEEE, 2013.

\bibitem{lecun2015deep}
Yann LeCun, Yoshua Bengio, and Geoffrey Hinton.
\newblock Deep learning.
\newblock {\em Nature}, 521(7553):436--444, 2015.

\bibitem{lee2010uncovering}
Kyumin Lee, James Caverlee, and Steve Webb.
\newblock Uncovering social spammers: social honeypots+ machine learning.
\newblock In {\em SIGIR'10}.

\bibitem{lesce1990scan}
Tony Lesce.
\newblock Scan: Deception detection by scientific content analysis.
\newblock {\em Law and Order}, 38(8):3--6, 1990.

\bibitem{li2016survey}
Yaliang Li, Jing Gao, Chuishi Meng, Qi~Li, Lu~Su, Bo~Zhao, Wei Fan, and Jiawei
  Han.
\newblock A survey on truth discovery.
\newblock {\em ACM Sigkdd Explorations Newsletter}, 17(2):1--16, 2016.

\bibitem{ling2003auc}
Charles~X Ling, Jin Huang, and Harry Zhang.
\newblock Auc: a statistically consistent and more discriminating measure than
  accuracy.

\bibitem{ma2016detecting}
Jing Ma, Wei Gao, Prasenjit Mitra, Sejeong Kwon, Bernard~J Jansen, Kam-Fai
  Wong, and Meeyoung Cha.
\newblock Detecting rumors from microblogs with recurrent neural networks.

\bibitem{ma2015detect}
Jing Ma, Wei Gao, Zhongyu Wei, Yueming Lu, and Kam-Fai Wong.
\newblock Detect rumors using time series of social context information on
  microblogging websites.
\newblock In {\em CIKM'15}.

\bibitem{magdy2010web}
Amr Magdy and Nayer Wanas.
\newblock Web-based statistical fact checking of textual documents.
\newblock In {\em Proceedings of the 2nd international workshop on Search and
  mining user-generated contents}, pages 103--110. ACM, 2010.

\bibitem{menczer2016spread}
Filippo Menczer.
\newblock The spread of misinformation in social media.
\newblock In {\em WWW'16}.

\bibitem{mitra2015credbank}
Tanushree Mitra and Eric Gilbert.
\newblock Credbank: A large-scale social media corpus with associated
  credibility annotations.
\newblock In {\em ICWSM'15}.

\bibitem{mohammad2017stance}
Saif~M Mohammad, Parinaz Sobhani, and Svetlana Kiritchenko.
\newblock Stance and sentiment in tweets.
\newblock {\em ACM Transactions on Internet Technology (TOIT)}, 17(3):26, 2017.

\bibitem{morstatter2016can}
Fred Morstatter, Harsh Dani, Justin Sampson, and Huan Liu.
\newblock Can one tamper with the sample api?: Toward neutralizing bias from
  spam and bot content.
\newblock In {\em WWW'16}.

\bibitem{morstatter2016new}
Fred Morstatter, Liang Wu, Tahora~H Nazer, Kathleen~M Carley, and Huan Liu.
\newblock A new approach to bot detection: Striking the balance between
  precision and recall.
\newblock In {\em ASONAM'16}.

\bibitem{mukherjee2015leveraging}
Subhabrata Mukherjee and Gerhard Weikum.
\newblock Leveraging joint interactions for credibility analysis in news
  communities.
\newblock In {\em CIKM'15}.

\bibitem{mustafaraj2017fake}
Eni Mustafaraj and Panagiotis~Takis Metaxas.
\newblock The fake news spreading plague: Was it preventable?
\newblock {\em arXiv preprint arXiv:1703.06988}, 2017.

\bibitem{nickerson1998confirmation}
Raymond~S Nickerson.
\newblock Confirmation bias: A ubiquitous phenomenon in many guises.
\newblock {\em Review of general psychology}, 2(2):175, 1998.

\bibitem{nyhan2010corrections}
Brendan Nyhan and Jason Reifler.
\newblock When corrections fail: The persistence of political misperceptions.
\newblock {\em Political Behavior}, 32(2):303--330, 2010.

\bibitem{paulrussian}
Christopher Paul and Miriam Matthews.
\newblock The russian “firehose of falsehood” propaganda model.

\bibitem{ping2013review}
Dong ping Tian et~al.
\newblock A review on image feature extraction and representation techniques.
\newblock {\em International Journal of Multimedia and Ubiquitous Engineering},
  8(4):385--396, 2013.

\bibitem{potthast2017stylometric}
Martin Potthast, Johannes Kiesel, Kevin Reinartz, Janek Bevendorff, and Benno
  Stein.
\newblock A stylometric inquiry into hyperpartisan and fake news.
\newblock {\em arXiv preprint arXiv:1702.05638}, 2017.

\bibitem{potthast2016clickbait}
Martin Potthast, Sebastian K{\"o}psel, Benno Stein, and Matthias Hagen.
\newblock Clickbait detection.
\newblock In {\em European Conference on Information Retrieval}, pages
  810--817. Springer, 2016.

\bibitem{qazvinian2011rumor}
Vahed Qazvinian, Emily Rosengren, Dragomir~R Radev, and Qiaozhu Mei.
\newblock Rumor has it: Identifying misinformation in microblogs.
\newblock In {\em EMNLP'11}.

\bibitem{quattrociocchi2016echo}
Walter Quattrociocchi, Antonio Scala, and Cass~R Sunstein.
\newblock Echo chambers on facebook.
\newblock 2016.

\bibitem{rubin2015deception}
Victoria~L Rubin, Yimin Chen, and Niall~J Conroy.
\newblock Deception detection for news: three types of fakes.
\newblock {\em Proceedings of the Association for Information Science and
  Technology}, 52(1):1--4, 2015.

\bibitem{rubin2016fake}
Victoria~L Rubin, Niall~J Conroy, Yimin Chen, and Sarah Cornwell.
\newblock Fake news or truth? using satirical cues to detect potentially
  misleading news.
\newblock In {\em Proceedings of NAACL-HLT}, pages 7--17, 2016.

\bibitem{rubin2015truth}
Victoria~L Rubin and Tatiana Lukoianova.
\newblock Truth and deception at the rhetorical structure level.
\newblock {\em Journal of the Association for Information Science and
  Technology}, 66(5):905--917, 2015.

\bibitem{ruchansky2017csi}
Natali Ruchansky, Sungyong Seo, and Yan Liu.
\newblock Csi: A hybrid deep model for fake news.
\newblock {\em arXiv preprint arXiv:1703.06959}, 2017.

\bibitem{sampson2016leveraging}
Justin Sampson, Fred Morstatter, Liang Wu, and Huan Liu.
\newblock Leveraging the implicit structure within social media for emergent
  rumor detection.
\newblock In {\em CIKM'15}.

\bibitem{shao2016hoaxy}
Chengcheng Shao, Giovanni~Luca Ciampaglia, Alessandro Flammini, and Filippo
  Menczer.
\newblock Hoaxy: A platform for tracking online misinformation.
\newblock In {\em WWW'16}.

\bibitem{shi2016fact}
Baoxu Shi and Tim Weninger.
\newblock Fact checking in heterogeneous information networks.
\newblock In {\em WWW'16}.

\bibitem{shu2017user}
Kai Shu, Suhang Wang, Jiliang Tang, Reza Zafarani, and Huan Liu.
\newblock User identity linkage across online social networks: A review.
\newblock {\em ACM SIGKDD Explorations Newsletter}, 18(2):5--17, 2017.

\bibitem{suwajanakorn2017synthesizing}
Supasorn Suwajanakorn, Steven~M Seitz, and Ira Kemelmacher-Shlizerman.
\newblock Synthesizing obama: learning lip sync from audio.
\newblock {\em ACM Transactions on Graphics (TOG)}, 36(4):95, 2017.

\bibitem{tacchini2017some}
Eugenio Tacchini, Gabriele Ballarin, Marco~L Della~Vedova, Stefano Moret, and
  Luca de~Alfaro.
\newblock Some like it hoax: Automated fake news detection in social networks.
\newblock {\em arXiv preprint arXiv:1704.07506}, 2017.

\bibitem{tajfel1979integrative}
Henri Tajfel and John~C Turner.
\newblock An integrative theory of intergroup conflict.
\newblock {\em The social psychology of intergroup relations}, 33(47):74, 1979.

\bibitem{tajfel2004social}
Henri Tajfel and John~C Turner.
\newblock The social identity theory of intergroup behavior.
\newblock 2004.

\bibitem{tang2015line}
Jian Tang, Meng Qu, Mingzhe Wang, Ming Zhang, Jun Yan, and Qiaozhu Mei.
\newblock Line: Large-scale information network embedding.
\newblock In {\em WWW'15}.

\bibitem{tang2014mining}
Jiliang Tang, Yi~Chang, and Huan Liu.
\newblock Mining social media with social theories: a survey.
\newblock {\em ACM SIGKDD Explorations Newsletter}, 15(2):20--29, 2014.

\bibitem{thies2016face2face}
Justus Thies, Michael Zollhofer, Marc Stamminger, Christian Theobalt, and
  Matthias Nie{\ss}ner.
\newblock Face2face: Real-time face capture and reenactment of rgb videos.
\newblock In {\em CVPR'16}.

\bibitem{tversky1992advances}
Amos Tversky and Daniel Kahneman.
\newblock Advances in prospect theory: Cumulative representation of
  uncertainty.
\newblock {\em Journal of Risk and uncertainty}, 5(4):297--323, 1992.

\bibitem{undeutsch1967beurteilung}
Udo Undeutsch.
\newblock Beurteilung der glaubhaftigkeit von aussagen.
\newblock {\em Handbuch der psychologie}, 11:26--181, 1967.

\bibitem{vlachos2014fact}
Andreas Vlachos and Sebastian Riedel.
\newblock Fact checking: Task definition and dataset construction.
\newblock {\em ACL'14}.

\bibitem{vrij2005criteria}
Aldert Vrij.
\newblock Criteria-based content analysis: A qualitative review of the first 37
  studies.
\newblock {\em Psychology, Public Policy, and Law}, 11(1):3, 2005.

\bibitem{wang2017attributed}
Suhang Wang, Charu Aggarwal, Jiliang Tang, and Huan Liu.
\newblock Attributed signed network embedding.
\newblock In {\em CIKM'17}.

\bibitem{wang2017signed}
Suhang Wang, Jiliang Tang, Charu Aggarwal, Yi~Chang, and Huan Liu.
\newblock Signed network embedding in social media.
\newblock In {\em SDM'17}.

\bibitem{wang2016linked}
Suhang Wang, Jiliang Tang, Charu Aggarwal, and Huan Liu.
\newblock Linked document embedding for classification.
\newblock In {\em CIKM'16}.

\bibitem{wang2016paired}
Suhang Wang, Jiliang Tang, Fred Morstatter, and Huan Liu.
\newblock Paired restricted boltzmann machine for linked data.
\newblock In {\em CIKM'16}.

\bibitem{wang2017your}
Suhang Wang, Yilin Wang, Jiliang Tang, Kai Shu, Suhas Ranganath, and Huan Liu.
\newblock What your images reveal: Exploiting visual contents for
  point-of-interest recommendation.
\newblock In {\em WWW'17}.

\bibitem{wang2017liar}
William~Yang Wang.
\newblock " liar, liar pants on fire": A new benchmark dataset for fake news
  detection.
\newblock {\em arXiv preprint arXiv:1705.00648}, 2017.

\bibitem{wang2015unsupervised}
Yilin Wang, Suhang Wang, Jiliang Tang, Huan Liu, and Baoxin Li.
\newblock Unsupervised sentiment analysis for social media images.
\newblock In {\em IJCAI}, pages 2378--2379, 2015.

\bibitem{ward1997naive}
Andrew Ward, L~Ross, E~Reed, E~Turiel, and T~Brown.
\newblock Naive realism in everyday life: Implications for social conflict and
  misunderstanding.
\newblock {\em Values and knowledge}, pages 103--135, 1997.

\bibitem{weikum2017computers}
Gerhard Weikum.
\newblock What computers should know, shouldn't know, and shouldn't believe.
\newblock In {\em WWW'17}.

\bibitem{wu2016chapter}
L~Wu, F~Morstatter, X~Hu, and H~Liu.
\newblock Chapter 5: Mining misinformation in social media, 2016.

\bibitem{wu2017adaptive}
Liang Wu, Xia Hu, Fred Morstatter, and Huan Liu.
\newblock Adaptive spammer detection with sparse group modeling.
\newblock In {\em ICWSM'17}.

\bibitem{wu2017gleaning}
Liang Wu, Jundong Li, Xia Hu, and Huan Liu.
\newblock Gleaning wisdom from the past: Early detection of emerging rumors in
  social media.
\newblock In {\em SDM'17}.

\bibitem{wu2016mining}
Liang Wu, Fred Morstatter, Xia Hu, and Huan Liu.
\newblock Mining misinformation in social media.
\newblock {\em Big Data in Complex and Social Networks}, pages 123--152, 2016.

\bibitem{wu2014toward}
You Wu, Pankaj~K Agarwal, Chengkai Li, Jun Yang, and Cong Yu.
\newblock Toward computational fact-checking.
\newblock {\em Proceedings of the VLDB Endowment}, 7(7):589--600, 2014.

\bibitem{yang2012automatic}
Fan Yang, Yang Liu, Xiaohui Yu, and Min Yang.
\newblock Automatic detection of rumor on sina weibo.
\newblock In {\em Proceedings of the ACM SIGKDD Workshop on Mining Data
  Semantics}, page~13. ACM, 2012.

\bibitem{zajonc1968attitudinal}
Robert~B Zajonc.
\newblock Attitudinal effects of mere exposure.
\newblock {\em Journal of personality and social psychology}, 9(2p2):1, 1968.

\bibitem{zajonc2001mere}
Robert~B Zajonc.
\newblock Mere exposure: A gateway to the subliminal.
\newblock {\em Current directions in psychological science}, 10(6):224--228,
  2001.

\bibitem{zubiaga2017detection}
Arkaitz Zubiaga, Ahmet Aker, Kalina Bontcheva, Maria Liakata, and Rob Procter.
\newblock Detection and resolution of rumours in social media: A survey.
\newblock {\em arXiv preprint arXiv:1704.00656}, 2017.

\end{thebibliography}

\end{document}